\documentclass[11pt,preprint]{aastex}
\usepackage{epsfig}

\newcommand{\msun}{\rm M_{\odot}}

\def\gs{\mathrel{\raise0.35ex\hbox{$\scriptstyle >$}\kern-0.6em
\lower0.40ex\hbox{{$\scriptstyle \sim$}}}}
\def\ls{\mathrel{\raise0.35ex\hbox{$\scriptstyle <$}\kern-0.6em
\lower0.40ex\hbox{{$\scriptstyle \sim$}}}}

\shorttitle{Deviations from the S-K relations}

\begin{document}

\title{Deviations from the Schmidt-Kennicutt relations during early galaxy evolution}

\author{Padelis P. Papadopoulos\altaffilmark{1},
Federico I. Pelupessy\altaffilmark{2}}
\altaffiltext{1}{Argelander-Institut f\"ur Astronomie, Auf dem H\"ugel 71, D-53121 Bonn, Germany}
\altaffiltext{2}{Leiden Observatory, Leiden University,
   PO Box 9513, 2300 RA Leiden, The Netherlands}

\begin{abstract}
  
  We utilize detailed time-varying  models of the coupled evolution of
  stars  and the  HI, $\rm  H_2$, and  CO-bright H$_2$  gas  phases in
  galaxy-sized  numerical  simulations  to  explore the  evolution  of
  gas-rich and/or  metal-poor systems, expected to be  numerous in the
  Early Universe.  The inclusion of the CO-bright H$_2$ gas phase, and
  the  realistic rendering  of  star formation  as an  H$_2$-regulated
  process (and  the new feedback  processes that this  entails) allows
  the most realistic tracking  of strongly evolving galaxies, and much
  better comparison  with observations.   We find that  while galaxies
  eventually settle into  states conforming to Schmidt-Kennicutt (S-K)
  relations,  significant  and  systematic  deviations of  their  star
  formation rates (SFRs) from  the latter occur, especially pronounced
  and prolonged  for metal-poor systems.  The  largest such deviations
  occur  for gas-rich  galaxies during  early evolutionary  stages but
  also  during brief  periods at  later stages.   Given  that gas-rich
  and/or metal-poor  states of present-epoch galaxies  are expected in
  the Early Universe while a  much larger number of mergers frequently
  resets non-isolated  systems to gas-rich states,  even brief periods
  of sustained deviations  of their SFRs from those  expected from S-K
  relations  may come  to  characterize significant  periods of  their
  stellar   mass  built-up.    This   indicates  potentially   serious
  limitations of (S-K)-type relations as reliable sub-grid elements of
  star formation physics in  simulations of structure formation in the
  Early Universe.  We anticipate  that galaxies with marked deviations
  from the S-K  relations will be found at  high redshifts as unbiased
  inventories  of total  gas mass  become possible  with ALMA  and the
  EVLA.

\end{abstract}


\section{Introduction\label{sec:intro}}

Since it was first proposed  as a phenomenological relation linking HI
gas  mass and  star  formation surface  density  in galaxies  (Schmidt
1959),  and subsequently  better constrained  and re-formulated  as to
include  also   CO-bright  H$_2$  gas  (Kennicutt   1998,  2008),  the
Schmidt-Kennicutt  (hereafter S-K) relation:  $\rm \Sigma_{SFR}\propto
[\Sigma   (HI)]^k$   (k$\sim   $1-2),   has  provided   the   standard
observational framework relating the star formation rate (as a surface
density rate $\rm \Sigma_{SFR}$) to the gas supply in galaxies.  It is
also  an important  element  of the  sub-grid  star formation  physics
incorporated in galaxy evolution  and structure formation models (e.g.
Baugh et  al.  2005  and references therein;  Springel, Di  Matteo, \&
Hernquist  2005,  Schaye  \&   Della  Vecchia  2008)  where  numerical
resolution  limitations preclude  a  more detailed  treatment of  star
formation over the scales involved.  Many theoretical (Dopita \& Ryder
1994; Robertson  \& Kravtsov 2008),  and observational (e.g.   Wong \&
Blitz 2002; Bigiel et al.  2008) studies have been made to demonstrate
its  validity,  with the  most  important  recent  advances being  the
identification of the CO-bright H$_2$ gas as better correlated to star
formation in galaxies  than atomic hydrogen (Wong \&  Blitz 2002), and
the   direct   star-formation   role   of  its   dense   phase   ($\rm
n(H_2)>10^4\,cm^{-3}$) with a k$\sim $1 (Gao \& Solomon~2004).

Unfortunately past analytical and  numerical investigations of the S-K
relation did not include a  multi-phase ISM (though see Gerritsen 1997
for  an early  investigation  that includes  it)  or assumed  sub-grid
models reacting instantaneously to changes  in the global state of the
ISM. Thus they are ill-suited  to explore very gas-rich systems and/or
early  galaxy evolutionary  stages, when  the various  ISM  phases and
their  interplay  with  the   stellar  content  have  not  established
equilibrium.   Finally, such  models  cannot be  compared directly  to
observations since they do not include the H$_2$ gas phase (the direct
fuel of  star formation),  or do so  but assume  that all such  gas is
CO-bright (e.g.   Gnedin et  al. 2009).  The  latter is not  the case,
especially in a metal-poor and/or far-UV intense ISM environment (e.g.
Israel 1997; Maloney \& Black 1988; Pak et al.  1998), which is common
during  the  gas-rich  and  vigorously star-forming  epochs  in  early
galaxy~evolution.

\subsection{ISM+stars galaxy models: features and limitations }

Here, we use  our time-varying models of the  coupled evolution of HI,
H$_2$ gas  phases and stars  in galaxy-sized numerical  simulations to
investigate  the  emergence  and  possible  deviations  from  the  S-K
relations.   Full details  and tests  of our  method can  be  found in
Pelupessy,  Papadopoulos, \&  van  der Werf  (2006)  and Pelupessy  \&
Papadopoulos (2009).  We use an N-body/SPH code and solve for the full
thermodynamic evolution of  the WNM and CNM HI  phases (see Wolfire et
al. 2003)  assuming neither an equillibrium nor  an Effective Equation
of  State  (EOS), unlike  most  current  cosmological or  galaxy-sized
structure formation  models (e.g.  Springel \& Hernquist  2003; Cox et
al.  2006a,b; Narayanan et al.   2009).  It is worth pointing out that
in  such models  the  coldest ISM  phase  tracked is  usually at  $\rm
T_{kin}$$\sim $10$^4$\,K (i.e.  thermodynamically far removed from the
one  truly  forming the  stars),  and  a Schmidt-Kennicutt  (S-K)-type
relation  between that  phase and  star formation  rate  is postulated
(e.g.   Kravtsov et  al.  2004;  Governato et  al.  2007).   In recent
models  colder  gas  ($\sim  $300\,K) is  tracked,  but  instantaneous
equillibrium gas  thermodynamics remains an assumption  (e.g Tasker \&
Bryan 2008) while the molecular  gas phase and the feedback effects of
star formation on it are not  included (e.g.  Tasker \& Tan 2009).  In
our approach star formation is controlled by gravitational instability
via a  Jeans mass criterion, the  thermal state of the  gas is tracked
explicitely, while an H$_2$-richness  criterion for star formation can
be  applied in  addition  to that  of  gravitational instability  (see
Pelupessy \& Papadopoulos 2009 for details).

The code tracks the H$_2$ phase with a physical model for substructure
using a minimal set of  assumptions. It follows the H$_2$ formation on
dust grains and its  thermal \& far-UV induced destruction, accounting
for self-shielding  and dust shielding.  The CO-bright  H$_2$ phase is
indentified  as  a  post-processing  step, using  the  most  important
chemical reactions  (R{\"o}llig at al.  2006) solving  for the C$^{+}$
envelope per gas  cloud, as regulated by the  local far-UV field.  The
aspects  making  our  code  particularly suitable  for  examining  the
validity of  the S-K  relation over  a range of  conditions are:  a) a
time-varying  treatment  of   the  H$_2$$\leftrightarrow$HI  gas  mass
exchange  while tracking  the ISM  from Warm  Neutral Medium  (WNM) HI
($\rm T_k$$\sim  $10$^4$\,K, $\rm n\sim  $(0.1--1)\,cm$^{-3}$) to Cold
Neutral Medium (CNM) HI and H$_2$ gas ($\rm T_{k}$$\sim $(30--200)\,K,
$\rm n$$\sim$(10--200)\,$\rm cm^{-3}$), b) the versatility of using an
H$_2$-regulating star  formation criterion,  {\it in addition}  to the
regular gravitational  instability criterion, and c)  CO formation and
destruction  (for this  specific  version of  our  code).  The  latter
allows  direct comparisons  with observations  since CO  line emission
rather than H$_2$ is the real observable in~galaxies.

The most  realistic models  employ our H$_2$-regulated  star formation
which uses the  local $\rm H_2$ gas mass fraction  as a star formation
regulator  in  the  dynamical  setting  of  an  evolving  galaxy.   We
implement this  molecular regulated (MR) star  formation by converting
only  the molecular  ($\rm  f_m$)  mass fraction  of  an unstable  gas
particle  to stars.   Unlike  our other  simulations  (where H$_2$  is
tracked  but plays  no  role in  star  formation) that  need an  adhoc
parameter  $\epsilon_{SF}$  designating the  local  gas mass  fraction
converted into stars, the MR  models contain a physical basis for this
parameter while retaining the original  requirement of SF gas as Jeans
unstable  (see Pelupessy  et  al  2006 for  more  details and  tests).
Finally here we  limit ourselves to systems with  $\rm M_{baryons} \le
10^{10}\,M_{\odot}$  where we can  maintain the  ability to  track the
full  range  of  gas  densities  and temperatures  necessary  for  the
HI$\rightarrow $H$_2$ phase transition to take place (typically in the
densest and coldest regions of CNM HI~gas).

\section{Emergent S-K relations and deviations}

Stationary galaxy  models show that  S-K relations appear to  hold for
variety of galaxy types and ISM conditions (e.g. Dopita \& Ryder 1994;
Robertson  \& Kravtsov  2008),  while early  dynamic modeling  (albeit
without H$_2$  included, and utilizing a much  simplified ISM picture)
shows  such  relations emerging  {\it  after}  a dynamic  equillibrium
between  ISM phases  (WNM  and  CNM HI)  and  star-formation has  been
established (Gerritsen \& Icke 1997).

\begin{table*}
 \caption[]{Overview  of  galaxy  model  parameters.  The  models  are
 labelled   with  letters   A-H  with   their   structural  parameters
 (e.g.   mass,  metallicity)   listed.   The   effects   of  different
 resolutions  are  discussed   elsewhere  (Pelupessy  \&  Papadopoulos
 2009). The  gas distributions  of models D1  and E1 consist  of equal
 mass exponential and extended  disks, the others have purely extended
 gas distributions.  $R_{\rm  gas}$ gives the extend of  the gas disk,
 $\Sigma_{\rm gas}$  the central gas surface density  , $R_{\rm disk}$
 the  exponential scale length  of the  stellar disk  and $\Sigma_{\rm
 disk}$ the stellar density in the center.
 
  } \centering
  \label{tab:models}
  \begin{tabular}{ c | r  r  r  r  r  r  r  r  r}
  \hline
  \hline
  Model & $M_{bary}$ & $M_{halo}$          & $f_{gas}$ & $m_{par}$ & $Z$ & $R_{\rm gas}$  & $\Sigma_{\rm gas}$ & $R_{\rm disk}$ & $\Sigma_{\rm disk}$ \\
        & $\msun$ & $\msun$                &           & $\msun$   &     & kpc  & $\msun/pc^2$ & kpc & $\msun/pc^2$ \\
  \hline
  A1    & $10^8$     & $2.3\times10^9$     & 0.5 & 200  & 0.2 & 1.5 & 30  & 0.2 & 200 \\
  B1    & $10^9$     & $2.3\times10^{10}$  & 0.2 & 500  & 0.2 & 4.2 & 16  & 0.5 & 450 \\
  C1    & $10^9$     & $2.3\times10^{10}$  & 0.2 & 500  & 1.0 & 4.2 & 16  & 0.5 & 450\\
  D1    & $10^{9}$   & $2.3\times10^{11}$  & .99 & 1000 & 1.0 & 4.5 & 280 & 0.5 & 5 \\
  E1    & $10^{9}$   & $2.3\times10^{11}$  & .99 & 1000 & 0.2 & 4.5 & 280 & 0.5 & 5 \\
  F1    & $10^{10}$  & $2.3\times10^{12}$  & .1  & 500  & 1.0 & 12  & 40  & 1.4 & 670 \\
  G1    & $10^{10}$  & $2.3\times10^{12}$  & 0.5 & 2500 & 0.2 & 12  & 200 & 1.4 & 370\\
  H1    & $10^{10}$  & $2.3\times10^{12}$  & 0.5 & 2500 & 1.0 &  12  & 200 & 1.4 & 370\\
  \hline
\end{tabular}
\end{table*}

\begin{figure*}
 \centering
 \epsscale{0.32}
 \caption{
 Time dependence of the SFR versus that expected from the S-K relation 
 and the instantanous gas content. Panels are labelled with the
 model number, ``MR'' indicates that the full H$_2$-regulated star
 formation was used. The solid lines depict the actual 
 total star formation rate, the other lines give the SFRs expected from
 various applications of the S-K relation (with slope n=1.4) using:
 total gas surface density (dashed), $\rm H_2$ gas surface density (dash-dotted), 
 and CO-bright $\rm H_2$ gas surface density (dotted).  The (S-K)-deduced
 SFR lines are normalized so that their average after 300~Myr
 matches the average star formation over the same period. The horizontal
 lines mark the SFRs of the Milky Way (dashed), and
 the Large Magelanic Cloud (dotted),  re-normalized to gas content of
 each model shown (e.g. $\rm SFR_{RN}$(LMC)=
 $\left[ \rm M_{gas}(model)/M_{gas}(LMC)\right]$$\times $SFR(LMC))  
 }
 \plotone{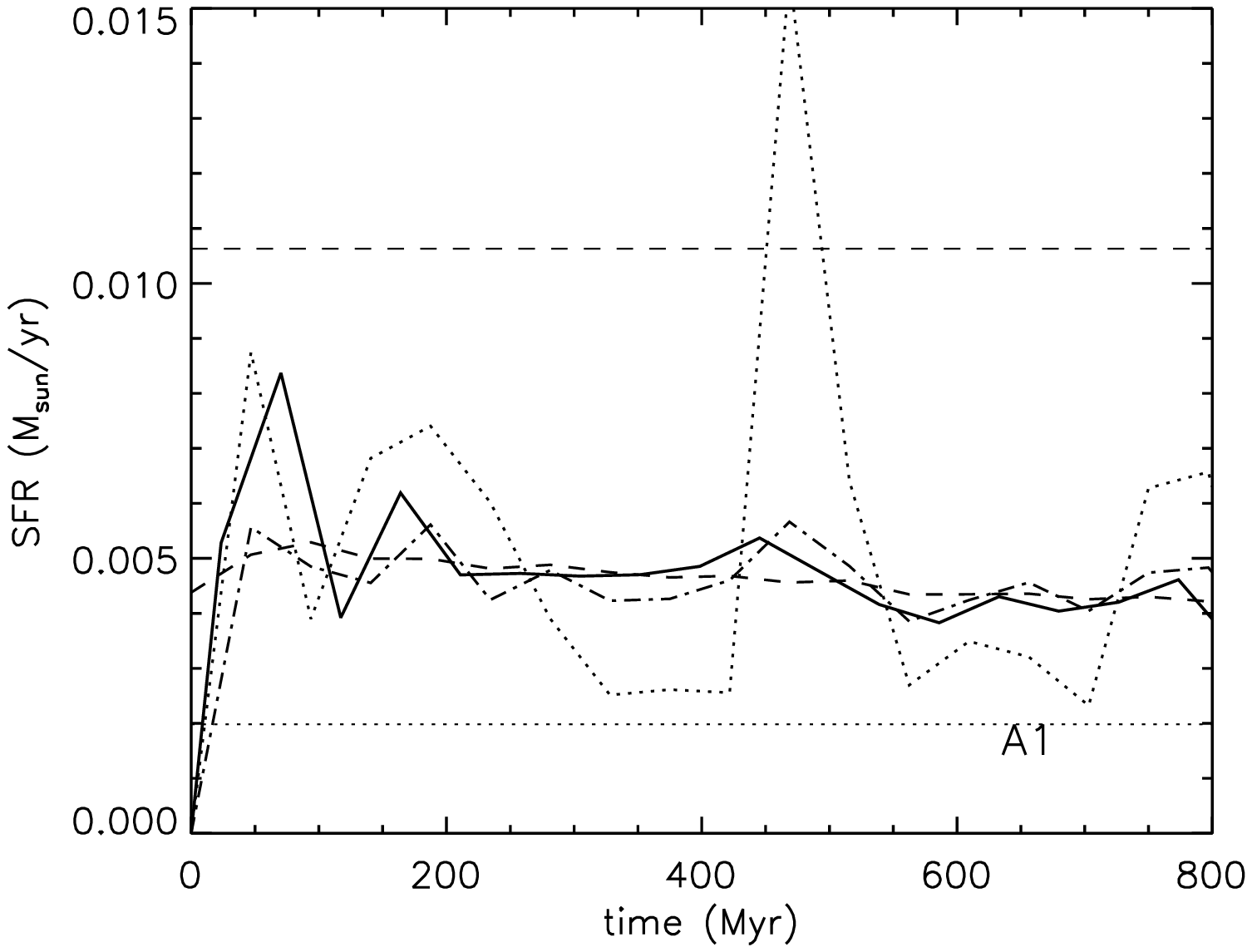}
 \plotone{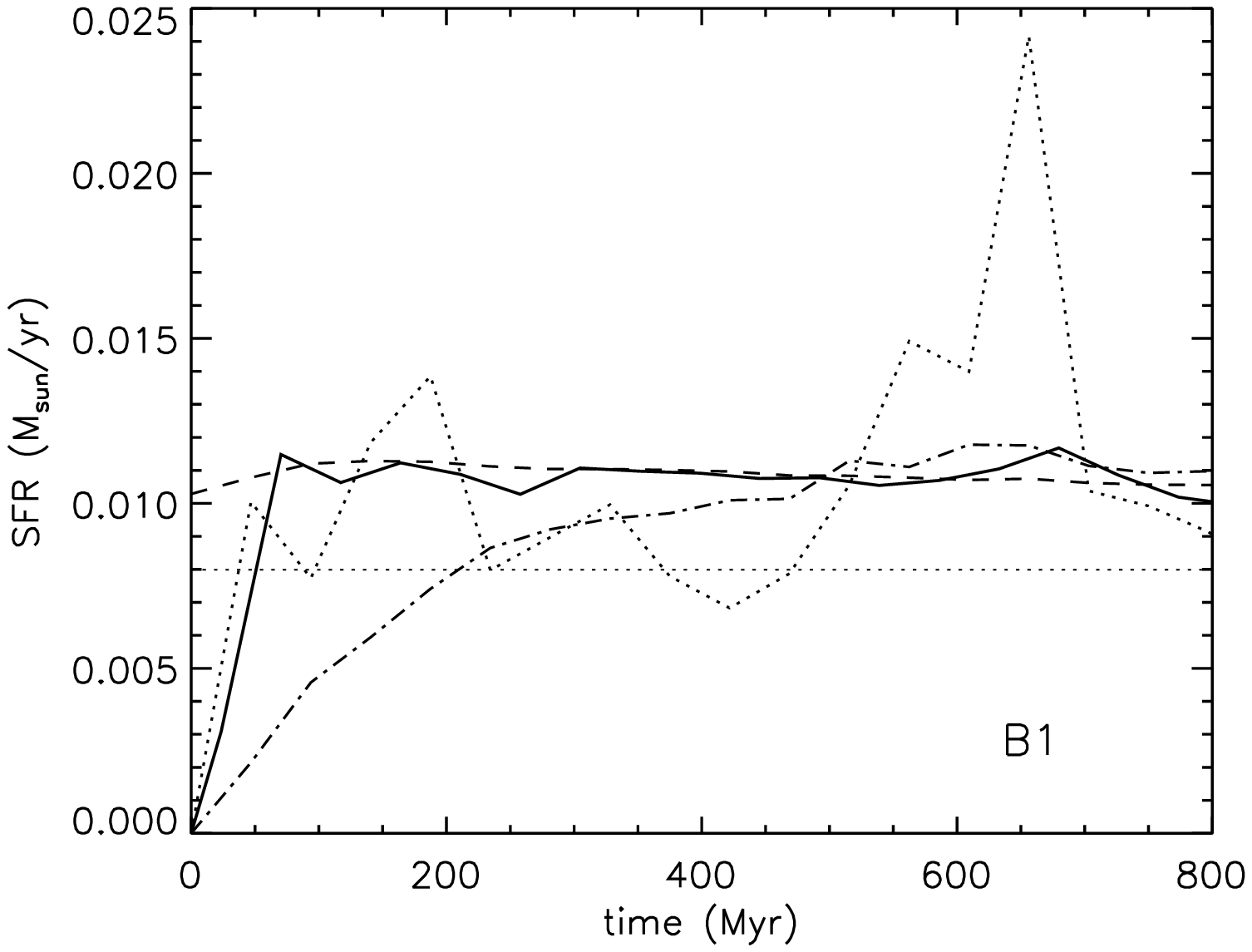}
 \plotone{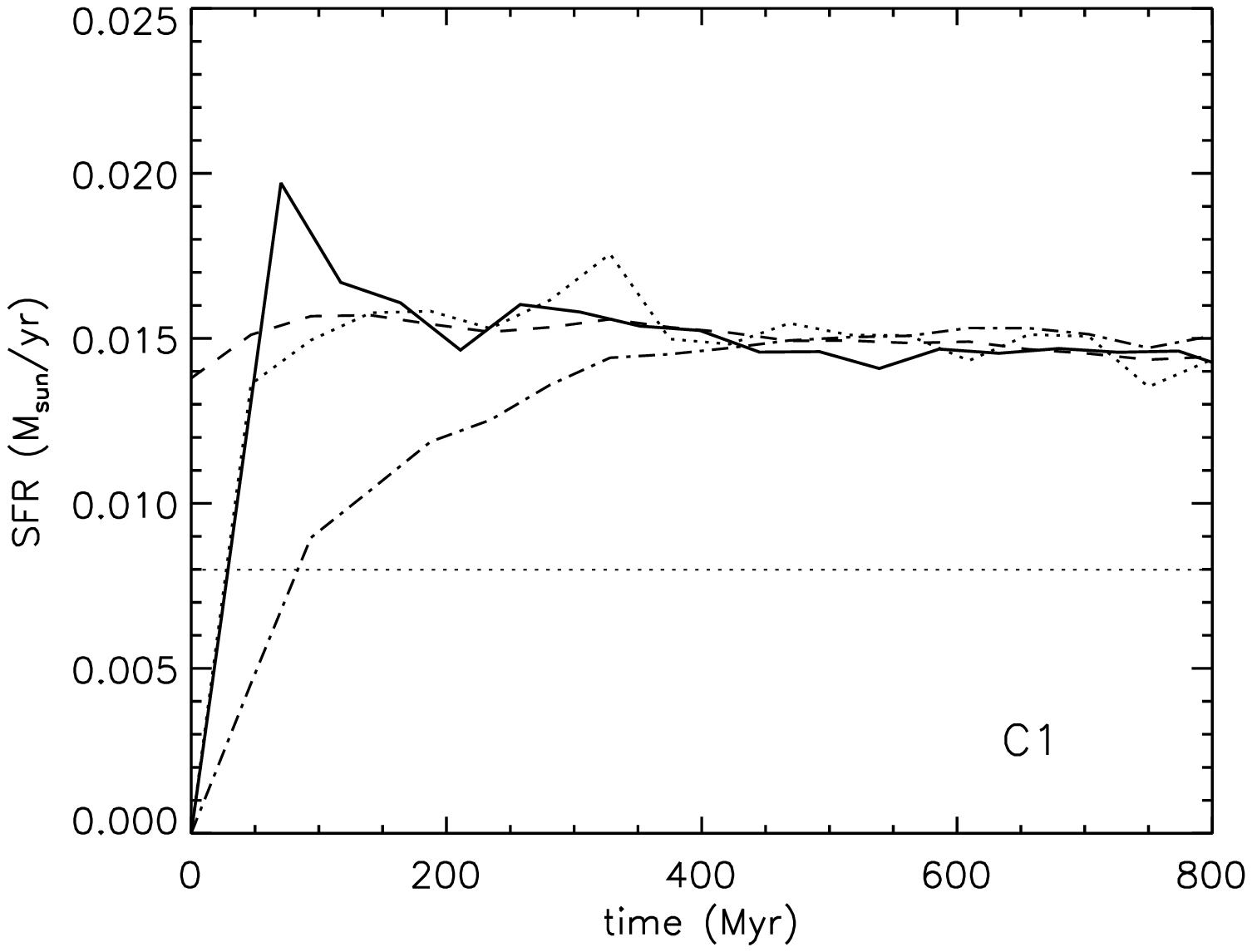}
 \plotone{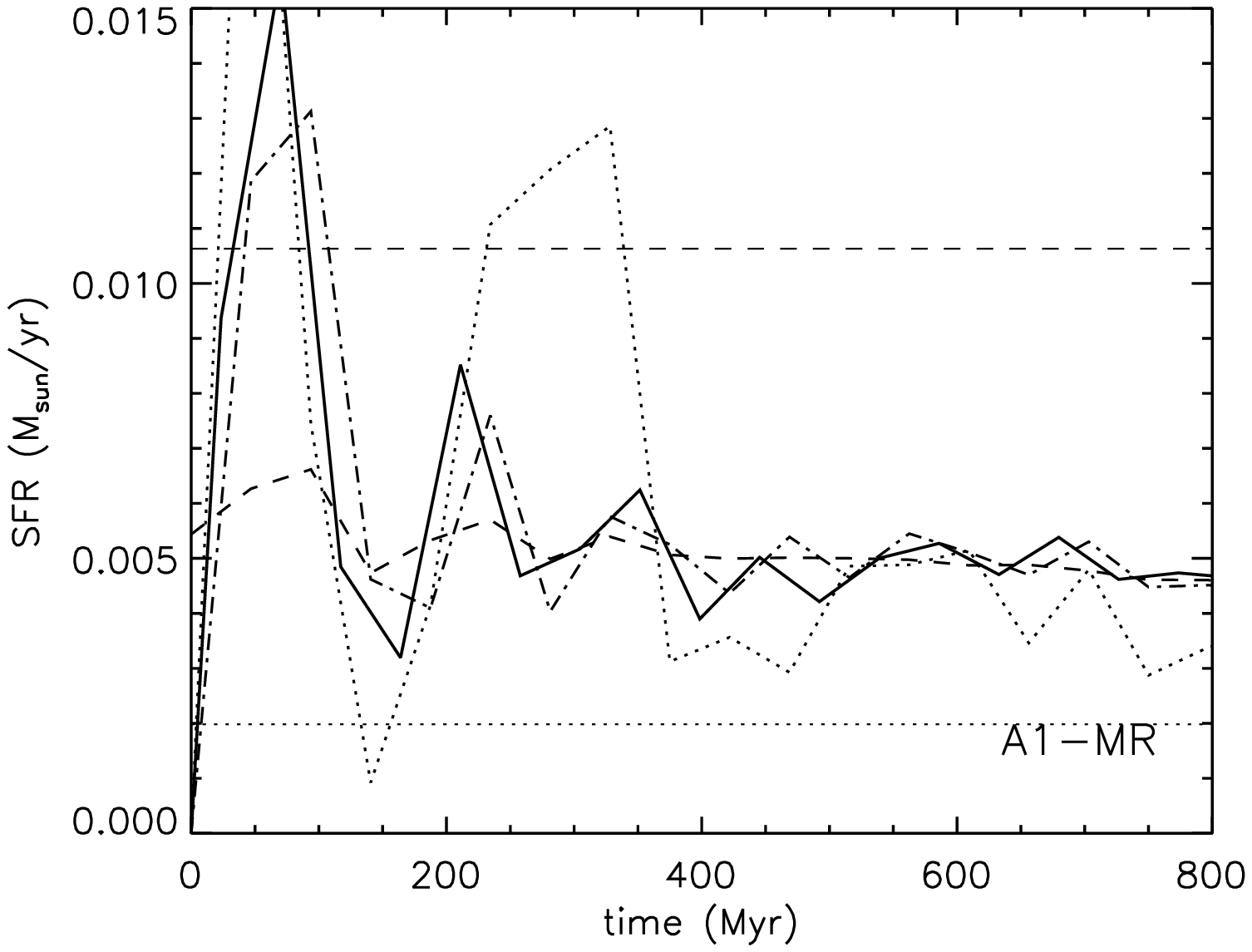}
 \plotone{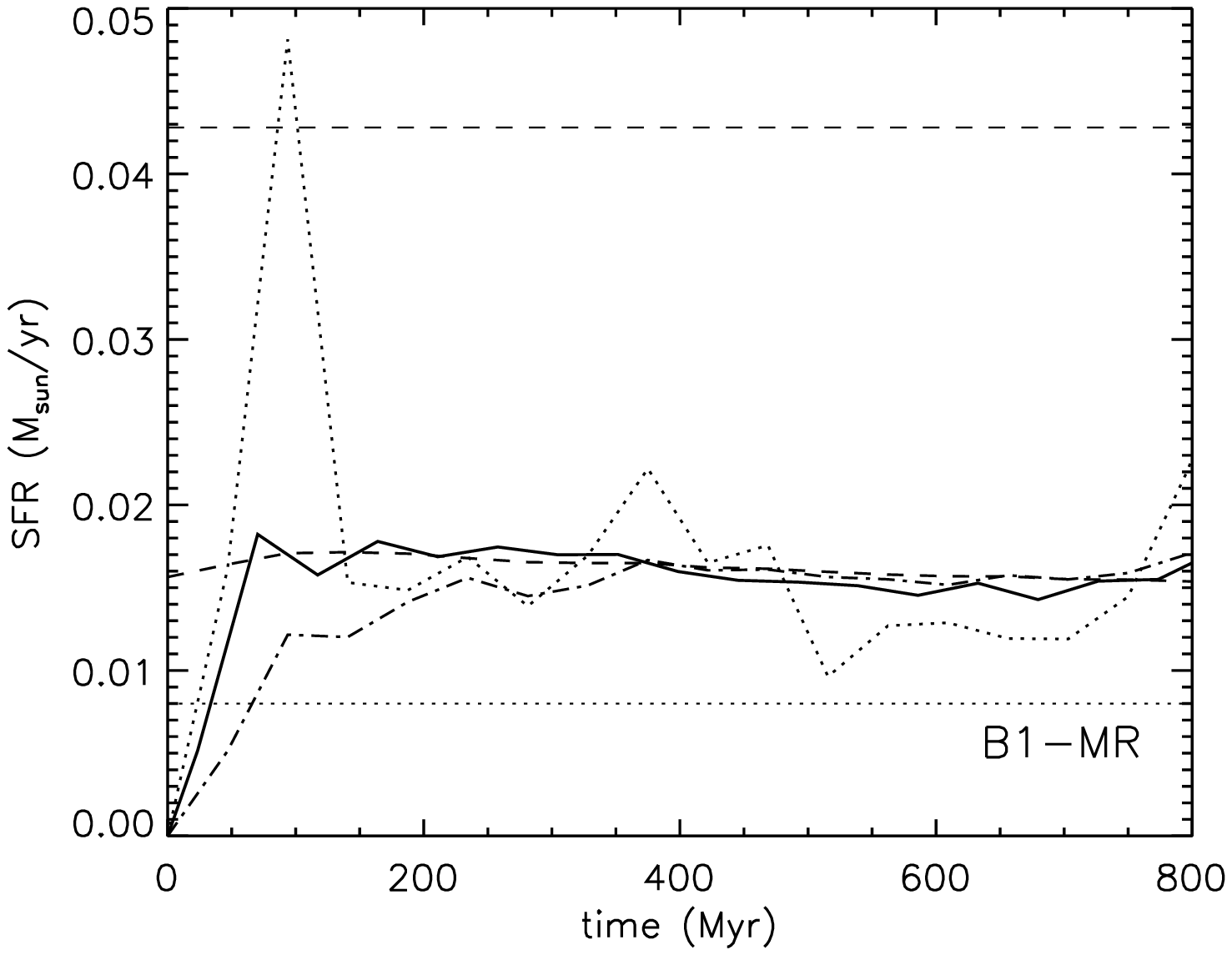}
 \plotone{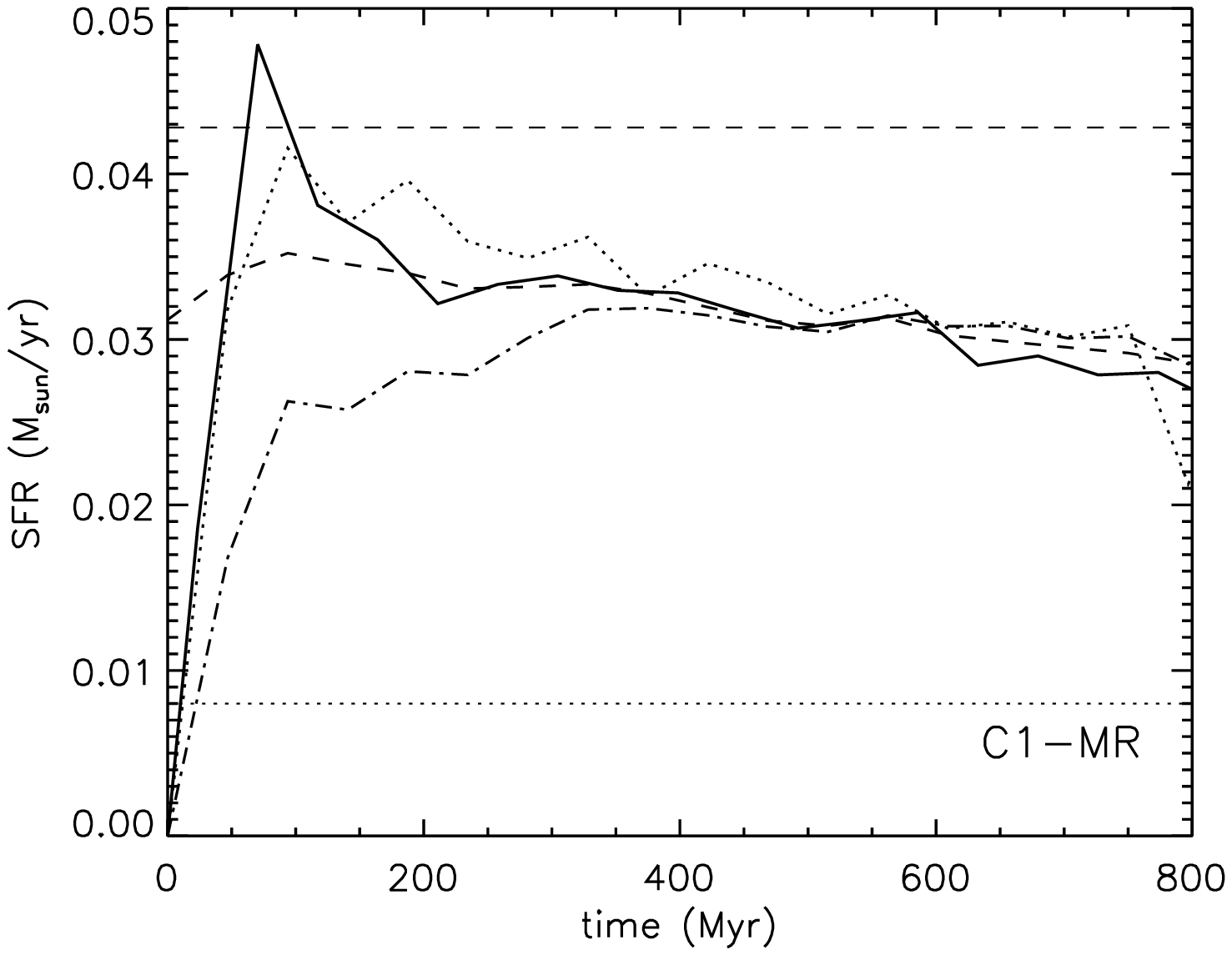}
 \plotone{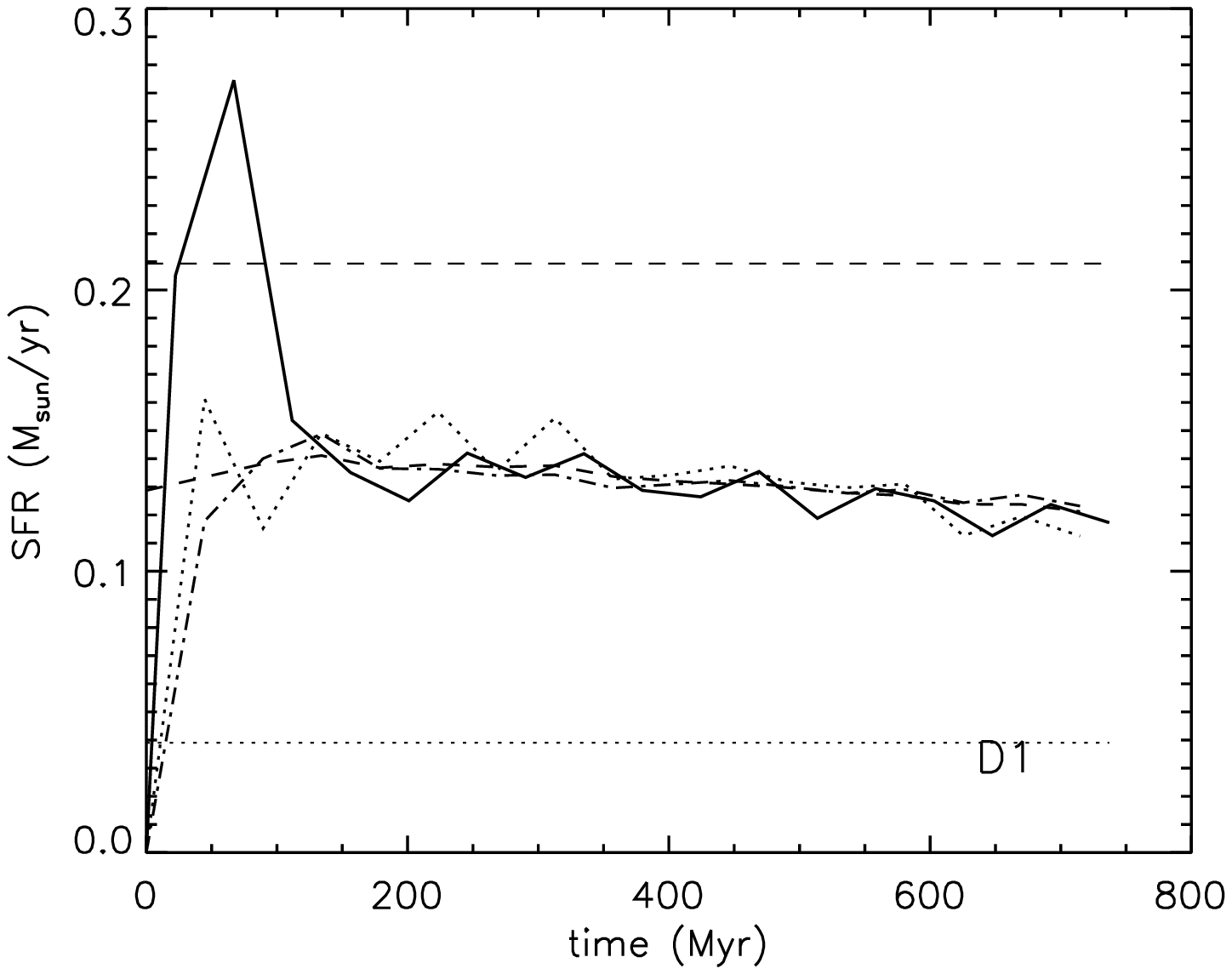}
 \plotone{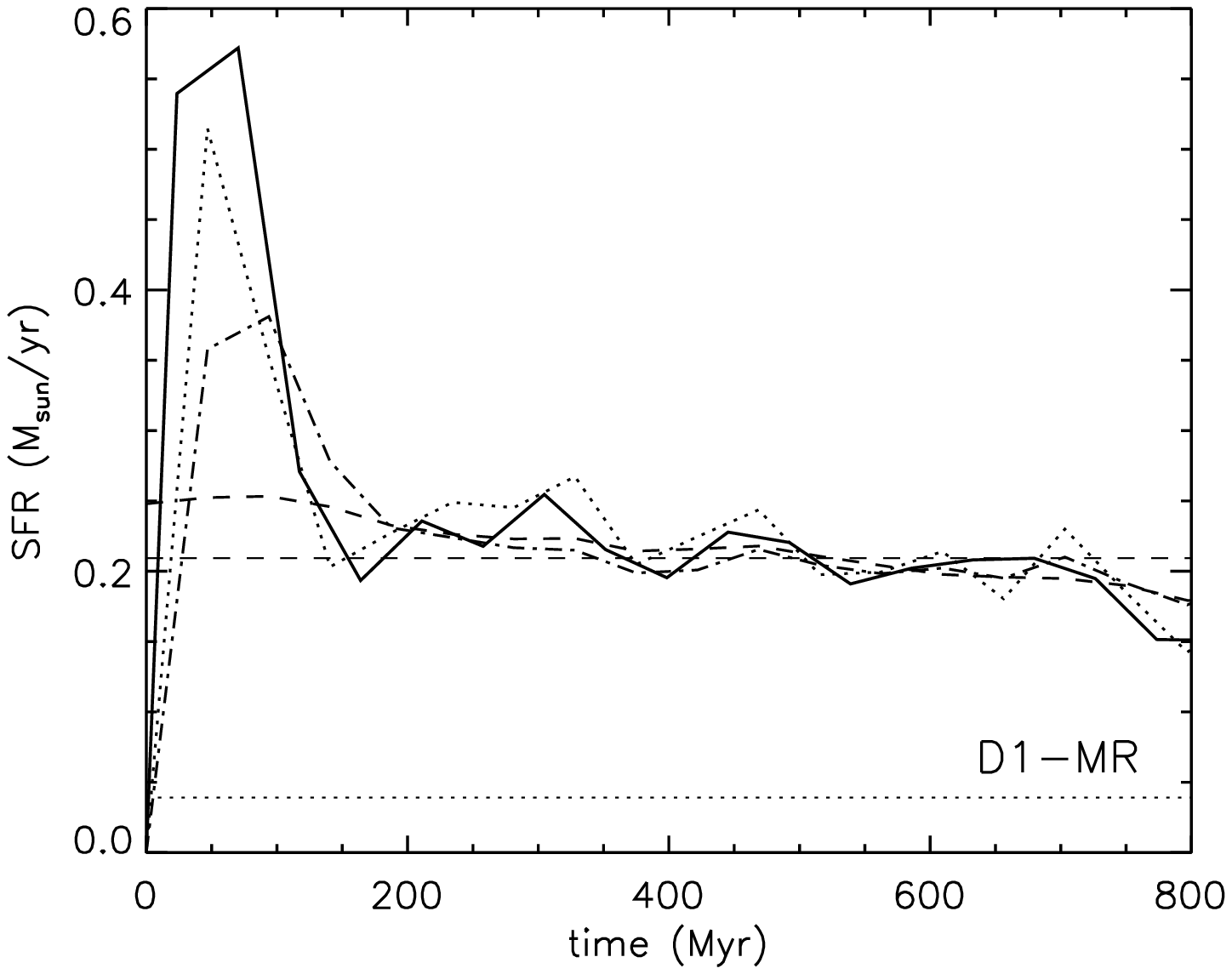}
 \plotone{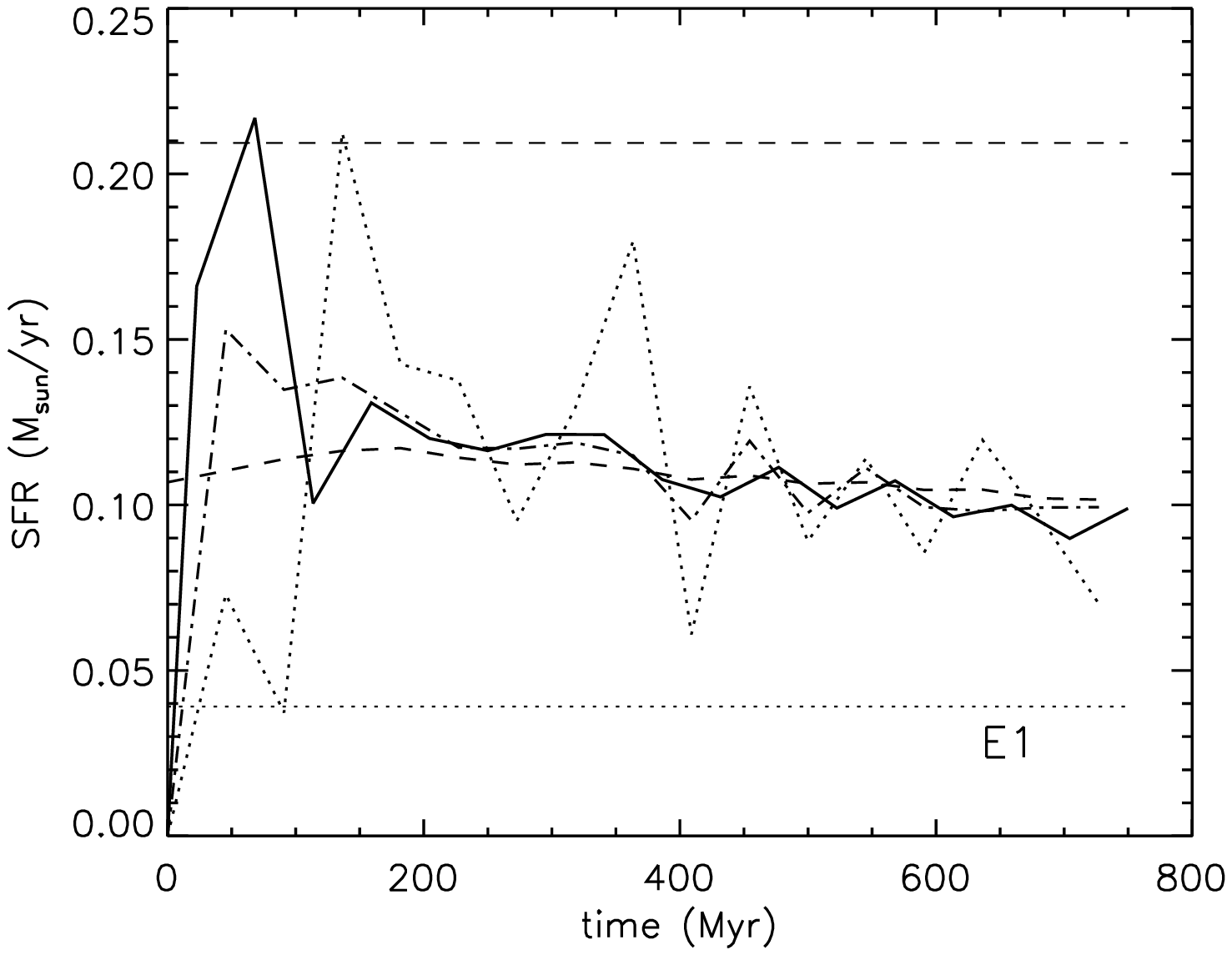}
 \label{fig:timeKS}
\end{figure*}

\begin{figure*}
 \centering
 \epsscale{0.32}
 \caption{A ``zoom'' into  the early galaxy evolution  for selected models in Figure 1, and
 results from the simulations of the large system F1 (lines and labels are  as in Figure 1).   
 For the two models  D1-MR and F1-MR where SFRs can be high, the re-normalized
 SFR of the Antennae galaxy, a nearby spectacular merger, is also
  shown (dash-dotted line at the top of the corresponding panels). 
}
 \plotone{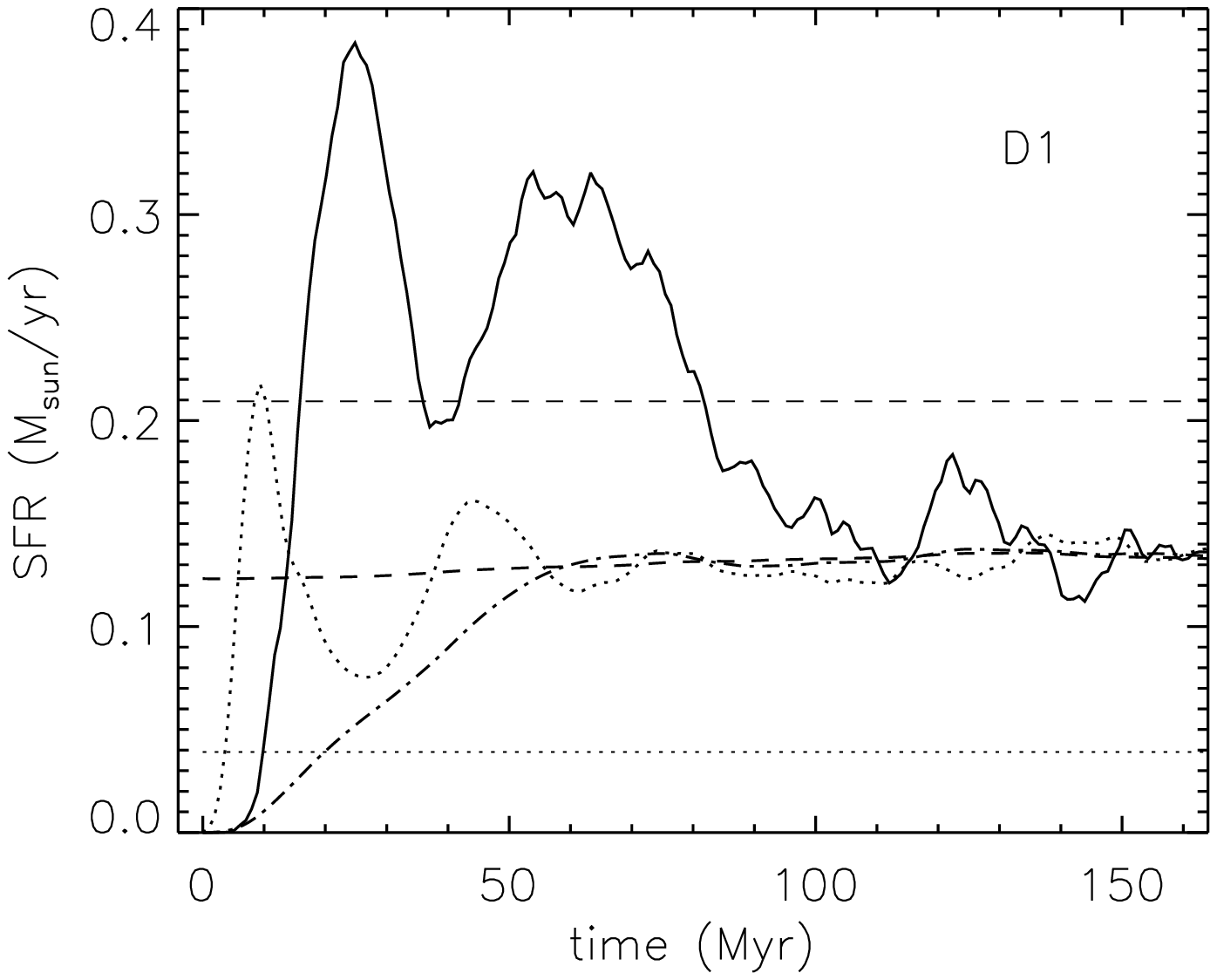}
 \plotone{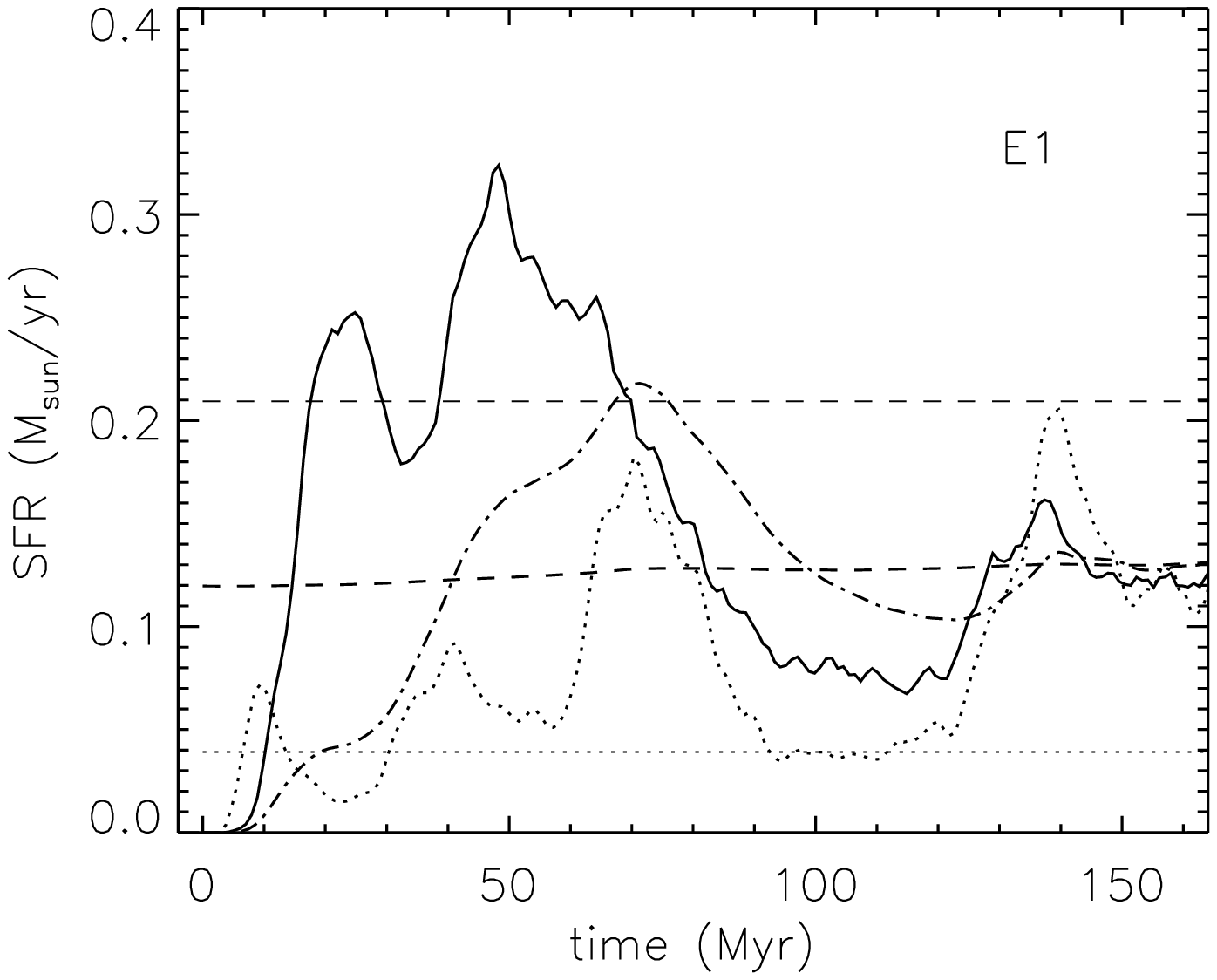}
 \plotone{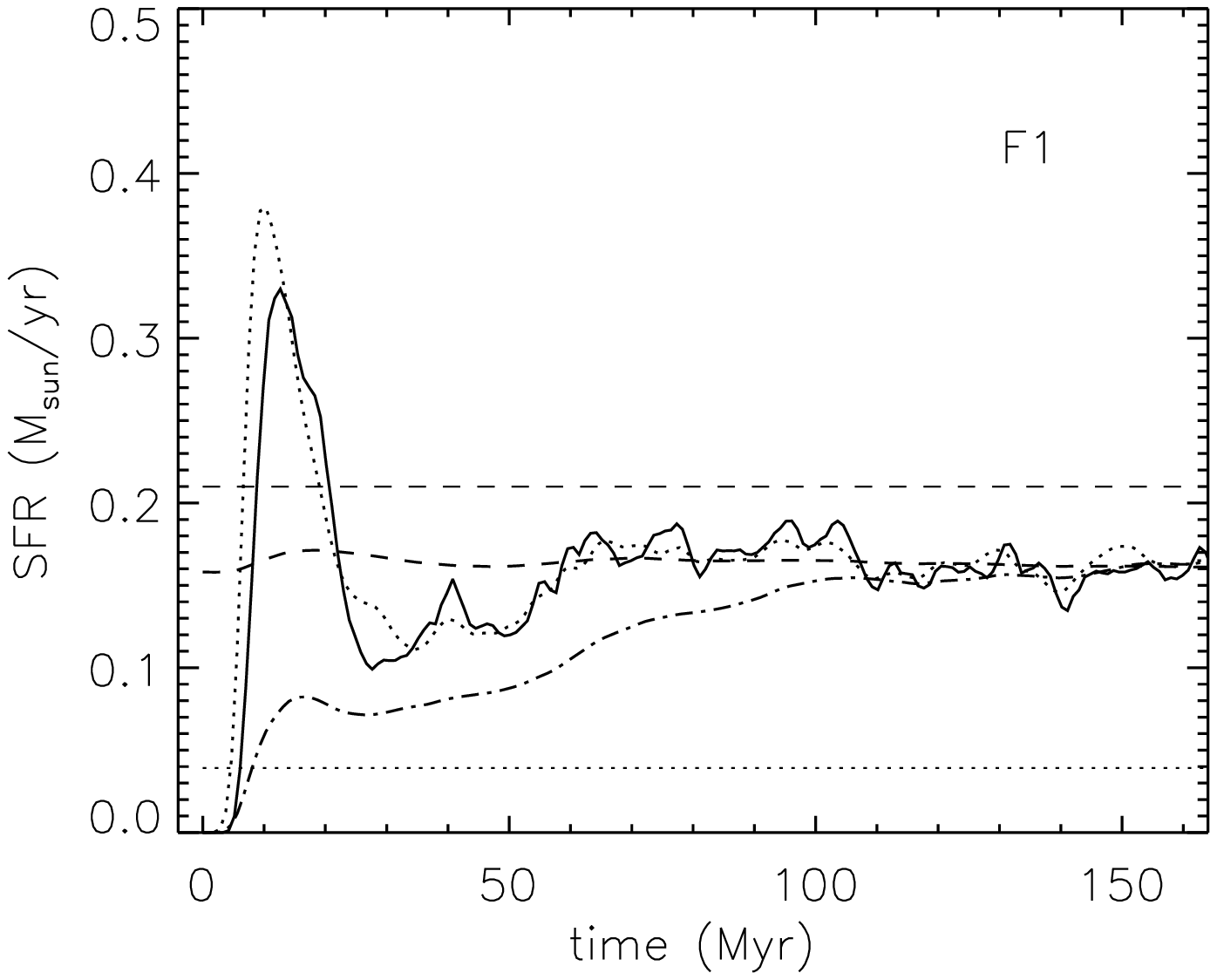}
 \plotone{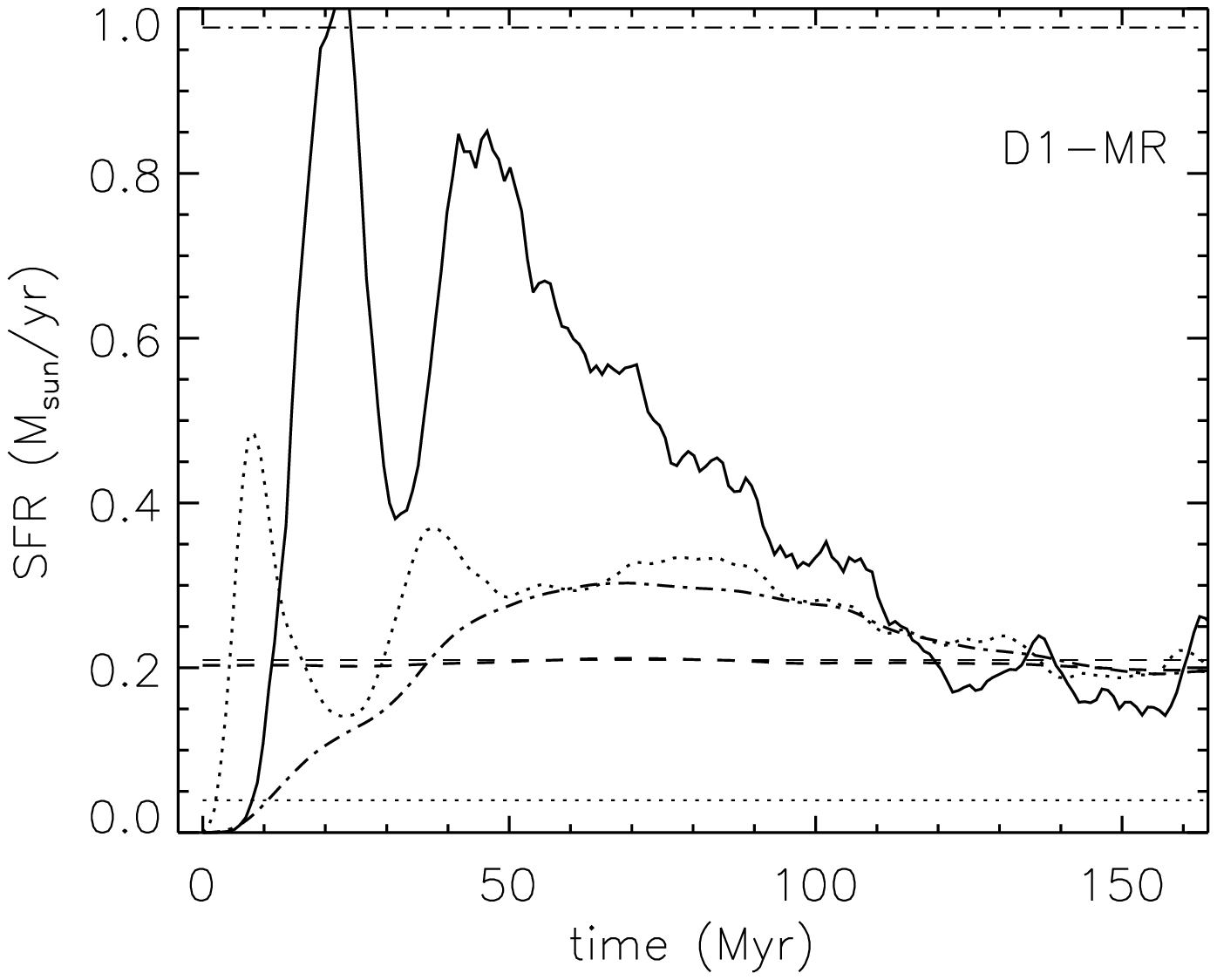}
 \plotone{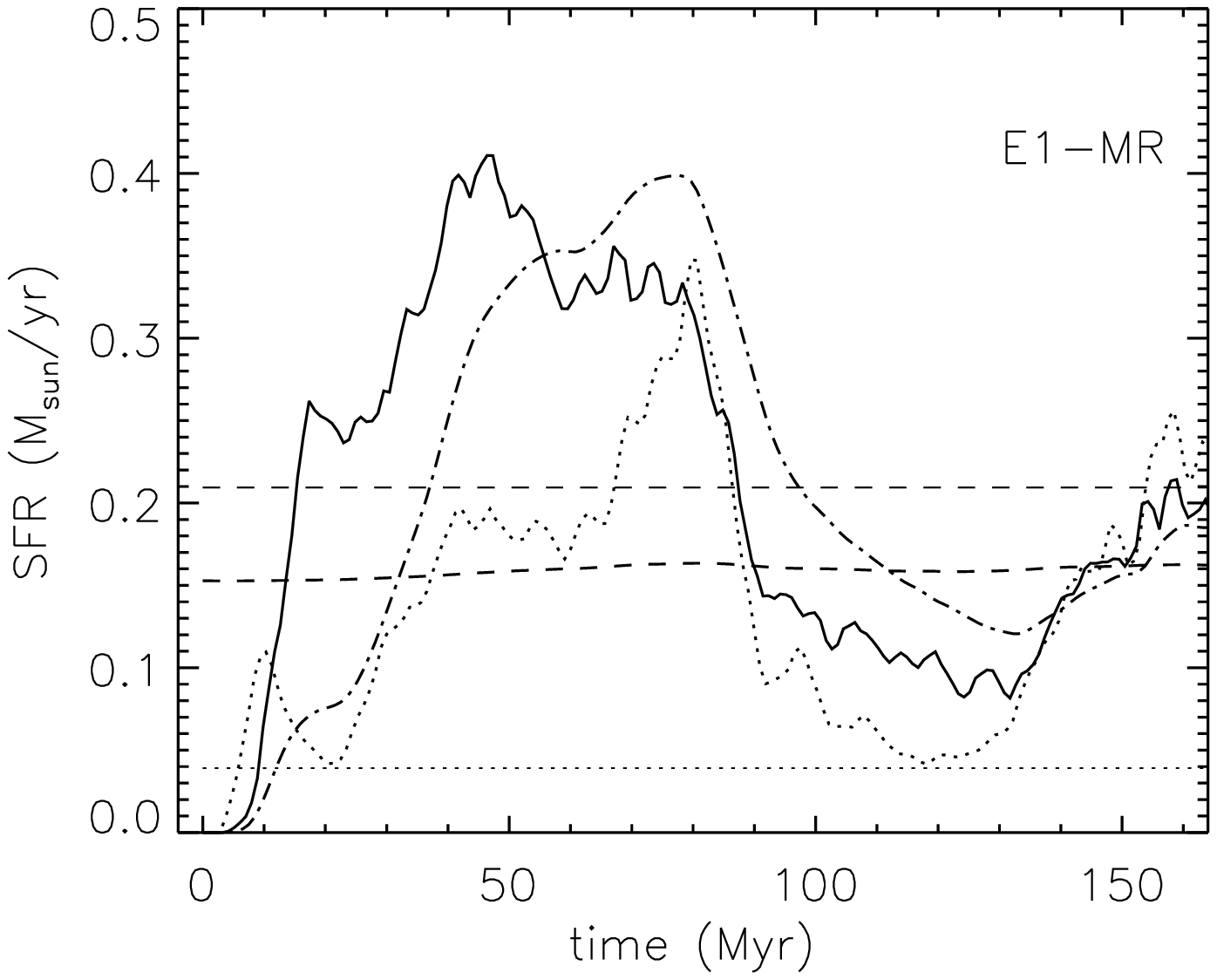}
 \plotone{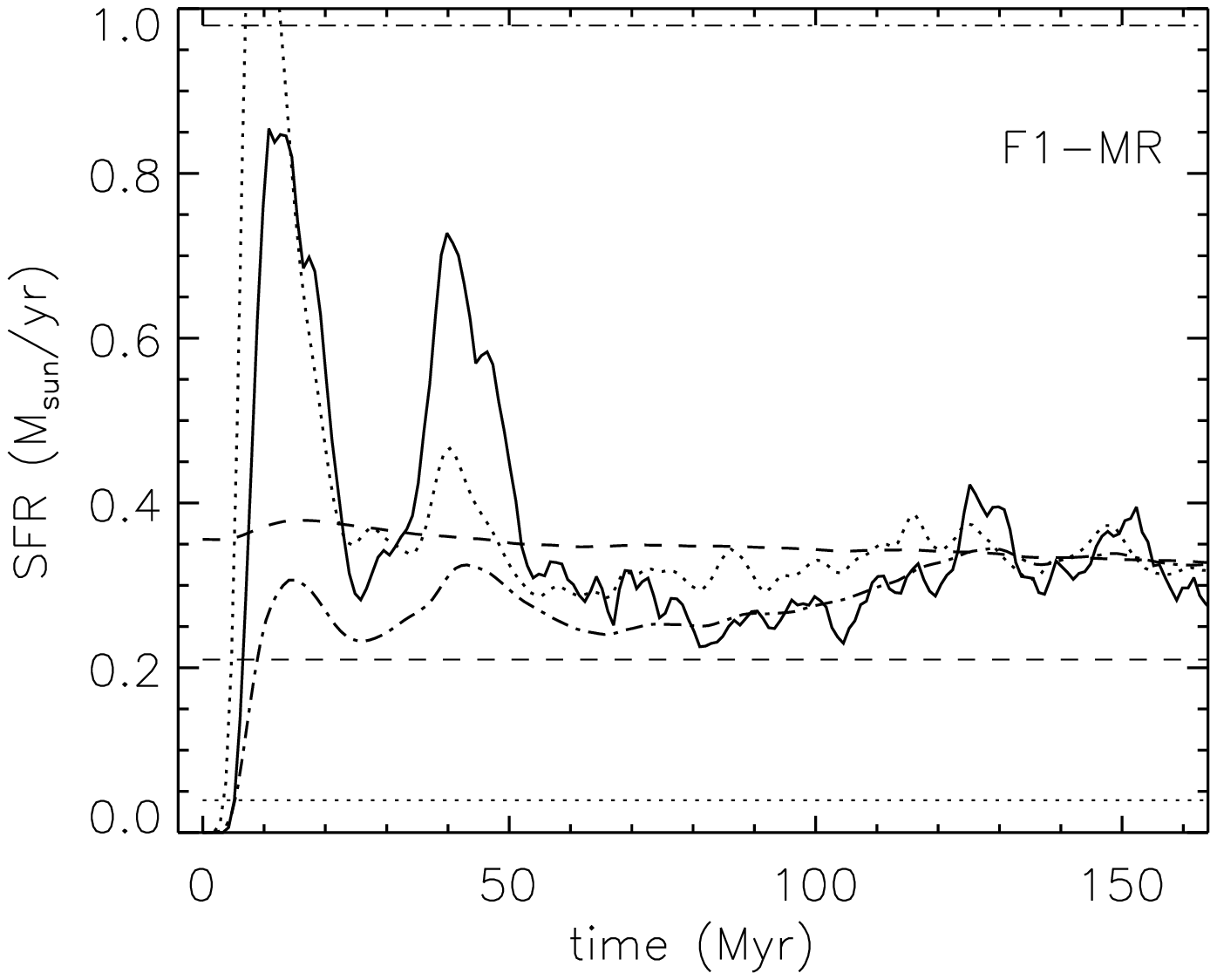}
 \label{fig:zoomKS}
\end{figure*}

\subsection{Model runs and Results}

The galaxy models we use  utilize the semi-analytic disc galaxy models
of  Mo, Mao  \&  White (1998)  and  Springel, Di  Matteo \&  Hernquist
(2005),  consisting of a  standard exponential  stellar and  a gaseous
component embedded  in a dark halo  with a Hernquist  profile (Table 1
gives an  overview of their properties).   The H$_2$ gas  mass and its
CO-bright  fractions  are  very   sensitive  to  the  metallicity  $Z$
(Pelupessy  \&  Papadopoulos  2009)  and  we thus  examine  models  at
Z=Z$_{\odot}$  and Z=Z$_{\odot}$/5.  Finally  we explore  systems with
gas  mass  fractions  ranging  from  those  typical  in  present-epoch
galaxies (10\%-20\%), to those found  recently in spiral disks at high
redshifts  (50\%), and even  up to  almost completely  gaseous systems
(99\%) representing the earliest stages in galaxy evolution.  The runs
start from gravitational and hydro-dynamical equilibrium (which is not
a thermodynamic, chemical or SF equilibrium) and all the gas initially
at  a WNM  HI phase.   Such initial  conditions can  be seen  as rough
proxies  for the  states expected  as  results of  (minor) mergers  or
infall from cold flows.  An initial WNM phase for the HI gas reservoir
is deemed  suitable for  diffuse, far-UV illuminated  gas of  the type
observed   at  high   Galactic   latitudes  and   expected  at   large
galactocentric  distances  of  spiral  galaxies (e.g.   Maloney  1993;
Kaufmann  et al.   2009).  The  models are  run for  $\approx $1\,Gyr,
after  which the resulting  gas and  star distributions  are extracted
and~mapped.

The central result we present here is that the models show significant
deviations from  the S-K relations, especially  the S-K(HI+H$_2$) one,
during strong galaxy evolution when  the ISM phases and star formation
are  out of  equillibrium.  This  is shown  in Figure~\ref{fig:timeKS}
from  where it can  be readily  discerned that  the SFRs  deviate from
those   anticipated  by   the  S-K   relations  especially   at  early
evolutionary  times, and such  deviations are  particularly pronounced
and prolonged for metal-poor systems  (A1, B1, and E1).  In the latter
cases  there  are  periods  when  the  CO-derived  S-K  relation  will
overestimate or underestimate the underlying star formation which, for
gas-rich {\it and}  metal-poor systems (E1), can last  well into later
evolution times (T$\sim $0.2--0.5\,Gyrs).   This seems to be an effect
of  the greater  sensitivity of  CO destruction  at  low metallicities
where  this molecule  survives only  in the  densest of  the  CNM gas,
itself spawing  star-forming regions very fast, which  in turn destroy
the CO.  The fact that this behavior emerges for both a small (A1) and
a  10$\times$ larger  metal-poor system  (B1) suggests  that  this SFR
``oscillation'' with  respect to a S-K(CO) relation  (Figure~1) is not
due  to  mere  stochastic  scatter  from  a  smaller  number  of  star
forming~sites.

Significant and systematic  deviations from the S-K(HI+H$_2$) relation
during early  galaxy evolution  epochs ($\rm T  $$\la $0.2--0.3\,Gyrs)
occur for most of our  models, and are particularly pronounced for the
very gas-rich systems (Fig.  1:  E1, D1 models).  The latter relation,
a common sub-grid element  of cosmological structure formation models,
thus  seems inapplicable  during periods  of strong  galaxy evolution.
Unfortunately the S-K(H$_2$) relation  may not fare much better during
such epochs (e.g.  C1, E1 in Fig.1, and F1 in Fig.2), and only S-K(CO)
remains a  good predictor  of the underlying  star formation  at early
evolutionary  times.  The  latter occurs  for metal-rich  systems with
moderate amounts  of gas (e.g.  C1,  C1-MR in Fig.1, and  F1, F1-MR in
Fig.2), i.e.   systems like those used  to derive the  S-K relation in
the local Universe.  The S-K(CO)  relation fares better since CO forms
only in  the dense and  cold regions of  the H$_2$ gas phase  and thus
tracks  the SF  sites more  closely,  but even  this {\it  observable}
relation seems to  underpredict the true SFR in  very gas-rich systems
(Fig.  1: D1,  E1).  This happens even for  the metal-rich system (D1)
where  CO tracks  the  $\rm  H_2$ distribution  well,  and thus  these
deviations  are not  due to  CO failing  to trace  the $\rm  H_2$ gas.
During  those  early  epochs  {\it  gas-rich  systems  can  appear  as
undergoing  periods of  very efficient  star formation}  (i.e.  little
CO-bright  $\rm  H_2$ gas  but  strong  ongoing  star formation),  and
application of  the S-K(CO) relation  using their {\it  observed} SFRs
would thus imply much more molecular gas than actually present.

Epochs during which  the actual SFRs can be  significantly {\it lower}
than the (S-K)-predicted  ones also exist (e.g. in  A1, A1-MR, B1, E1,
E1-MR),  and are  indeed expected  during intervals  of  strong galaxy
evolution  when  large, out-of-equillibrium,  amounts  of warm  ($\sim
$10$^3$--10$^4$\,K) HI  and H$_2$ gas  mass are produced  by spatially
extended and  almost coherent  starburst episodes.  During  such times
the star-forming CNM  H$_2$ gas phase may contain  little mass yet the
S-K(HI+H$_2$) relation cannot account for this as it considers all the
gas as star formation ``fuel'', even when in a phase thermodynamically
far-removed  from the  one  actually forming  stars  in galaxies.   In
metal-poor systems,  rather surprisingly,  we also find  periods where
even CO-bright gas experiences star formation lower than expected from
the  S-K(CO)   relation,  and  this  can  happen   even  during  later
evolutionary   times   (models:   A1,B1,E1).    A  recent   study   of
low-metallicity   high-redshift   systems   using   cosmological-sized
simulations also finds  marked deviations of the actual  SFR from that
expected from the S-K relations (Gnedin \& Kravtsov 2010) although its
applicability to actual observations is  hindered by the fact that the
CO  molecule (the true  observable) is  not considered.   Moreover the
larger volumes  modeled in  such studies necessitate  the use  of more
sub-grid  physics of  e.g.  H$_2$  formation, which  in turn  can make
their   galaxy-sized  results   more  dependant   on   the  particular
assumptions  made to  set up  the sub-grid  ISM model.

 Examining  the earliest  epochs  (T$<$150\,Myr) of  our most  massive
galaxy models at a finer timestep yields better resolved deviations of
their  intrinsic  SFRs  from  those  predicted by  the  S-K  relations
(Figures 2,  3).  During  those early times,  when the evolution  of a
galaxy is the  strongest, these deviations are the  largest and remain
most prominent  in very gas-rich  systems (e.g.  D1,~E1 in  Figure 2),
while they  persist for  longer times  in the metal  poor ones  (E1 in
Figure 1).  However it must be pointed out that significant deviations
can still  be found in galaxies  with lower gas  fractions, typical in
the  local Universe  (e.g.   models  A1, A1-MR,  B1,  B1-MR in  Figure
1). These are within the  significant dispersion of actual SFRs around
the S-K relation  observed in systems in the  local Universe ($\sim $a
factor of 10), and suggests  limitations of such relations as reliable
predictive tools of  star formation rate for a given  amount of gas to
better than an  order of magnitude, even for  normal galaxies.  If, as
our simulations suggest, such a dispersion is mostly non-stochastic in
nature and the result of  the various feedback factors in action (e.g.
SNR-induced    shocks,    far-UV    radiation    variations,    strong
non-equillibrium WNM$\leftrightarrow$CNM HI mass exchange) then use of
S-K relations as sub-grid elements of star formation physics in galaxy
formation  models  {\it  can   impart  serious  limitations  on  their
predictive power even for present-epoch disk galaxies.}

\subsection{Evolution of the largest gas-rich systems}

\begin{figure*}
\centering
\epsscale{1.1}
\plottwo{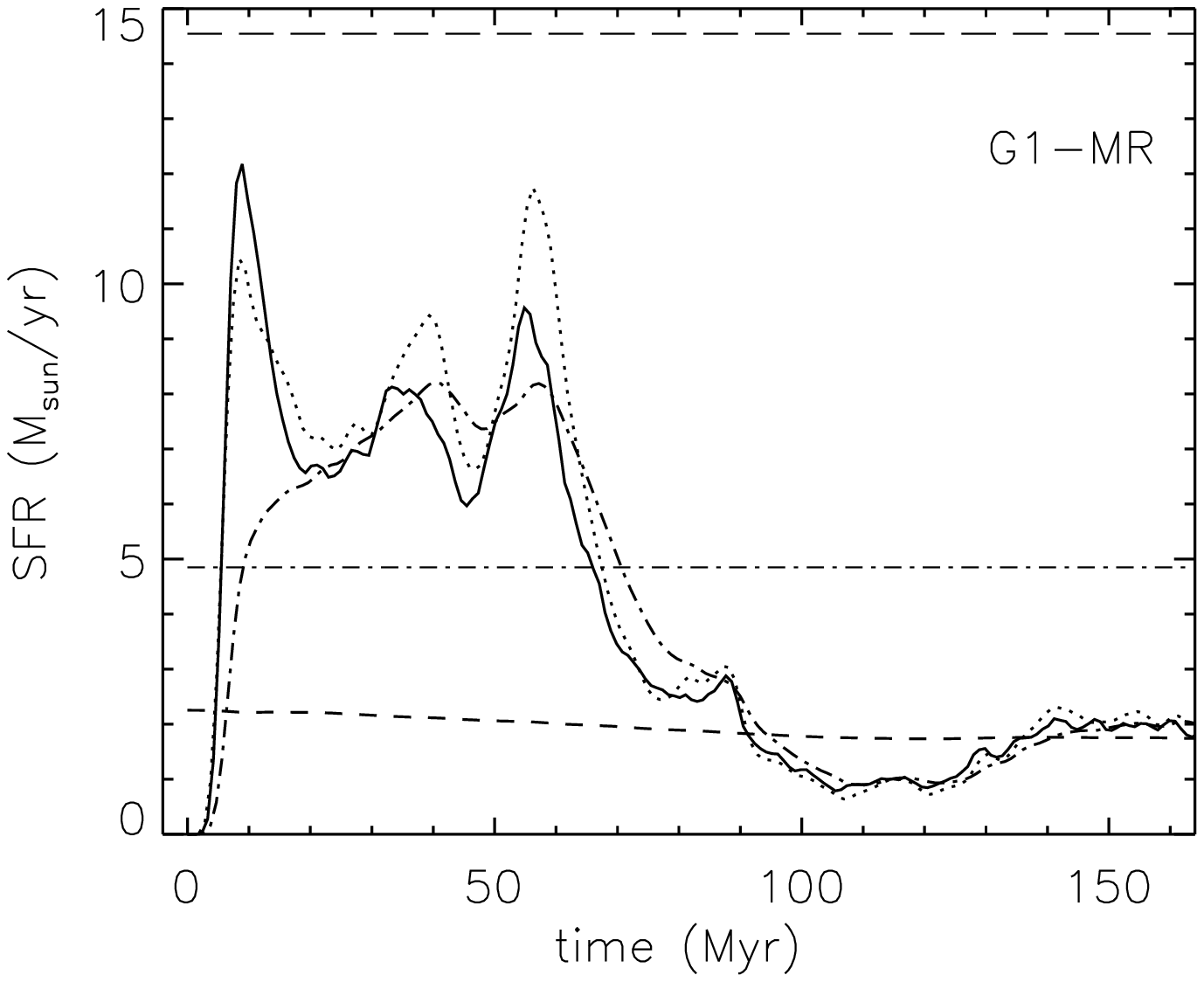}{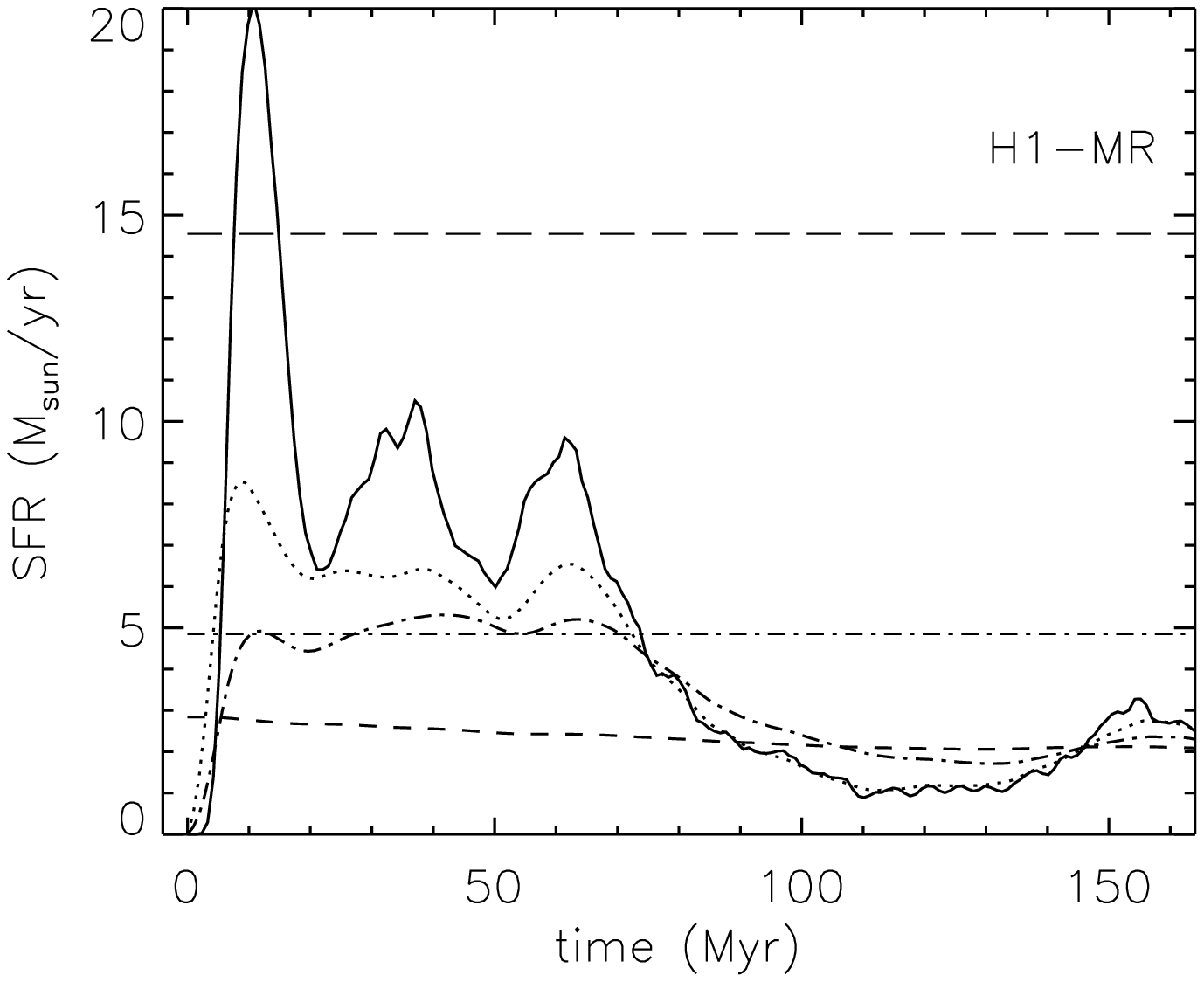}
\caption{The early evolution of the largest gas-rich systems with 50\%
gas mass  fraction and  metallicities of Z=0.2$\rm  Z_{\odot}$ (G1-MR)
and Z=$\rm Z_{\odot}$ (H1-MR) (see  Table 1 for details).  The SFRs from
the various S-K  relations are  denoted by lines as  in  Figures 1 and 2.
 Two horizontal lines now show  the SFR of the Antennae  galaxy (dash-dotted), 
and BzK galaxies (Daddi et al. 2010) (dashed), re-normalized to the gas
 content of our models.}
\end{figure*}

Interestingly the largest disk with the smallest gas mass fraction and
a solar metallicity (model F1)  settles relatively quickly into a full
conformance to the S-K relations  after $\sim $50\,Myr (Figure 2), and
this  is  the type  of  systems for  which  such  relations have  been
established in  the local  Universe.  Moreover unlike  previous cases,
the SFR  deduced from  the S-K(CO) relation  now remains close  to the
actual  one (Figure  2, models:  F1, F1-MR),  even during  those first
50\,Myr of  strong SFR time-dependance  and its large  deviations from
the  S-K(HI+H$_2$)  and S-K(H$_2$)  relations.   This underscores  the
central  role of the  CO-bright H$_2$  gas as  the phase  most closely
associated with  the star forming  sites, reflecting the fact  that CO
forms in the  densest regions of the CNM H$_2$ gas  which are also the
regions where star formation happens and proceeds the fastest.

 For disks with the same size  but gas mass fractions of 50\%, akin to
 galaxies found recently at high redshifts (Daddi et al. 2010; Tacconi
 et al.  2010),  we find the largest and  most sustained deviations of
 the  actual SFR  from  the S-K(HI+H$_2$)  relation  (Figure 3).   The
 S-K(H$_2$) and even  more so the S-K(CO) relation  follow SFR(t) more
 closely  though significant deviations  still exist.   Thus structure
 formation models where star formation of such disks is followed using
 an  S-K(HI+H$_2$) relation as  a sub-grid  element of  star formation
 physics may {\it significantly underestimate the rate of stellar mass
 built-up.}

\section{Discussion}
  
In the local Universe large variations of the so-called star formation
efficiency $\rm SFE=SFR/M_{gas}$ have been found in galaxies, with the
very  H$_2$-rich  Ultra  Luminous  Infrared Galaxies  (ULIRGs)  having
$10\times$ or higher SFEs than quiescent spirals.  If the framework of
the  S-K relation is  maintained such  variations imply  different (k)
exponents  for different galaxy  types, with  ULIRGs having  k$\sim $1
while quiescent spirals  and HI-rich objects reaching up  to k$\sim $2
(Schmidt 1959;  Wong \&  Blitz 2002; Gao  \& Solomon 2004).   This has
been  attributed to  a  strongly varying  fraction  of the  SF-fueling
versus  the total  gas reservoir  among galaxies,  with  the CO-bright
molecular gas and its even  denser HCN-bright phase as the actual star
forming  gas (Wong  \& Blitz  2003;  Gao \&  Solomon 2004;  Wu et  al.
2005).  Our numerical simulations corroborate this picture and further
reveal S-K  deviations as the  complex outcome of  strong, non-linear,
and highly non-equillibrium mass and energy exchange among the various
ISM phases  and the  stellar component during  times of  strong galaxy
evolution.  In that regard the  varying global SFE's and k-values then
simply  reflect  the ergodic  ``unfolding''  of such  non-equillibrium
events over sets  of galaxies in the local  Universe, and the attempts
to  parametrize  it  in  a  simple  fashion (i.e.   in  terms  of  S-K
relations).   The significant  dispersion that  is  typically observed
around  the S-K  relation (factors  of  $\sim $5-10  when derived  for
systems spanning a large range  of properties and SFRs) is then simply
masking  a deeper  set of  ongoing physical  processes in  the  ISM of
star-forming  galaxies.   As our  simulations  suggest, during  strong
galaxy  evolution  such processes  can  set  star formation  wandering
significantly above  or below a  ``mean'' S-K relation,  much reducing
its predictive power.

\subsection{Current observational evidence and biases}

Unfortunately unbiased studies of  the relation between star formation
and the  ambient gas supply are  currently possible only  in the local
Universe, with  an unbiased census of  the {\it total}  gas mass being
the main obstacle at high redshifts. This is because currently neither
HI  nor the  CO-bright H$_2$  gas mass  distribution can  be routinely
imaged  (via  the  21\,cm  and  CO  J=1--0  lines)  in  high  redshift
galaxies. Thus  a high-z study of  the S-K relation  using exactly the
same gas mass tracers used to  establish it locally is not possible at
present. As  our simulations indicate gas-rich systems  can stay close
to  the   S-K(CO)  relation,   but  deviate  substantially   from  the
S-K(HI+H$_2$) one.  Moreover,  since currently only CO J+1$\rightarrow
$J,  J+1$\geq $3  line observations  are typically  available  for the
distant Universe,  they introduce  {\it a bias  towards the  dense and
warm star-forming molecular gas} which in turn can lead to a seemingly
constant S-K(CO high-J) relation  across cosmic epoch (e.g. Tacconi et
al.  2010).  This may simply reflect the fact that the dense molecular
gas exciting such high-J CO  lines remains the direct ``fuel'' of star
formation with an almost constant  SF efficiency in all galaxies (e.g.
Gao \&  Solomon 2004) across  cosmic epoch, but leaves  open questions
regarding the S-K(HI+H$_2$) or  even the S-K(CO) relation (established
with the CO J=1--0 line) in gas-rich systems at high~redshifts.

  There  are some  early indications  for high-redshifts  systems with
much higher  SFRs for the amount  of CO-bright H$_2$  gas they contain
(e.g.  UV/optically selected galaxies,  Tacconi et al.  2008), as well
as evidence for systems  with large CO-bright molecular gas reservoirs
but with $\ga $10 times lower star-formation rates than those expected
from the S-K(CO) relation (Nesvadba  et al.  2009; Elbaz et al.  2009;
Dannerbauer et  al.  2009).  From our  study we expect  that once much
less biased gas mass measurements  in the distant Universe will become
possible  with ALMA  and the  EVLA, galaxies  systematically deviating
from (S-K)-type phenomenological relations will be uncovered, and such
deviations  will be  especially prominent  for metal-poor  and/or very
gas-rich star-forming systems.

\section{Conclusions} 

Our  results  are of  special  importance  for  the modeling  of  very
gas-rich and/or metal-poor progenitors of present-epoch galaxies found
in the  distant Universe,  or systems where  major gas  mass accretion
events frequently ``reset'' their evolutionary states back to gas-rich
ones.   In  such cases  the  non-equilibrium, non-linear,  mass/energy
exchange between the various ISM  phases and the stellar component may
come  to dominate significant  periods of  intense star  formation and
stellar mass built-up during which the S-K relation is not applicable.
In short  we find that it  may work well  for present-epoch metal-rich
spirals with  modest remaining gas  mass fractions (i.e.   systems for
which the (S-K) relations were  originally deduced), but not for their
very  gas-rich/metal-poor  progenitors  in  the Early  Universe.   The
importance of such deviations of  the actual star formation rates from
those  expected from  S-K  relations does  not  lie so  much in  their
magnitude (though  this must be explored further  for gas-rich systems
larger  than the  ones  modelled here)  rather  than their  systematic
nature  (e.g.   star-forming  systems  spending certain  periods  with
always higher or  always lower SFRs than those  estimated from the S-K
relations).   It is  the latter  that may  make  such phenomenological
relations poor  choices for the sub-grid physics  of star-formation in
rigorous structure formation models in a cosmological~setting.

Finally we note that when it comes to the large gas-rich galaxies that
are currently accessible at high redshifts our results, drawn for less
massive  systems, remain provisional.   Nevertheless for  more massive
and very  gas-rich star-forming systems  the larger amplitudes  of ISM
equilibrium-perturbing  agents (e.g.   SNs, far-UV  radiation fields),
and the shorter timescales  characterizing their variations, will more
likely  than not  exaggerate  the deviations  of  true star  formation
versus  the one expected  from (S-K)-type  phenomenological relations.
An  unbiased  observational effort  to  find  and study  (S-K)-deviant
galaxies at high redshifts (soon to be possible with ALMA and the EVLA
over a  wide range  of galaxy masses),  as well as  extending detailed
numerical  modeling  of  gas  and  stars towards  larger  systems  (as
computational capabilities improve), are important in establishing the
validity range of S-K relations and thus their utility as an important
sub-grid element in models of structure formation in the Universe.

\acknowledgments We would  like to take this opportunity  to thank the
referee for  several comments and suggestions that  improved this work
and expanded the scope of the original paper, especially when it comes
to the simulations of the larger systems presented here.

\bibliographystyle{astron}

\end{document}